\documentclass[12pt]{iopart}
\usepackage{hyperref}
\usepackage{graphicx}
\usepackage{color}
\usepackage{amssymb}
\usepackage[rmdefault=mdput,
            OMLmathbf         
           ]{isomath}
\usepackage{tikz}

\renewcommand{\vec}[1]{\mathbf{#1}}
\DeclareMathAlphabet\mathbfcal{OMS}{cmsy}{b}{n}

\begin{document}

\title{Berry phase of spin-one system in a rotating electric field}

\author{Abdaljalel Alizzi$^1$, Zurab K. Silagadze$^{1,2}$ and Artem Uskov$^1$}
\address{$^1$ Novosibirsk State University, 630 090, Novosibirsk, Russia}
\address{$^2$  Budker Institute of Nuclear Physics, 630 090, Novosibirsk,
Russia}
\ead{abdaljalel90@gmail.com}
\ead{silagadze@inp.nsk.su}
\ead{a.uskov+articles@alumni.nsu.ru}

\begin{abstract}
We consider in detail how the Berry phase arises in a rotating electric field in a model system with spin one. The goal is to help the student who first encountered this interesting problem, which is fraught with some subtleties that require attention in order not to go astray.

\noindent{\it Keywords\/}: Berry phase; Geometric phase; Time-dependent Hamiltonian.
\end{abstract}

\section{Introduction}
The geometric phase (Berry phase) is a simple and fundamental aspect of the adiabatic theorem in quantum mechanics \cite{Wilczek_1989,Bohm1993,Zwanziger_1990}. If $\psi(t)$ is a non-degenerate eigenstate of the time-dependent Hamiltonian $\hat H(t)$ at $t=0$ with an energy eigenvalue $E$, then the adiabatic theorem guarantees that as the Hamiltonian changes slowly, compared with characteristic time (the Planck constant divided by the energy gap between $E$ and the nearest energy level), then $\psi(t)$ will always remain the eigenvector of $\hat H(t)$ \cite{Sakurai_2011}. Therefore, if $\hat H(T)=\hat H(0)$, then $\psi(T)$ and $\psi(0)$ can differ only by a phase. In the seminal article \cite{Berry_1984}, Berry showed that this phase, in addition to the expected dynamic part $\alpha_d=-\frac{i}{\hbar}\int\limits_0^T E(\tau)d\tau$, contains also the geometric part $\alpha_g$, which is independent on dynamics and is determined exclusively by the geometry and topology of the space of parameters controlling the Hamiltonian.

Almost immediately it became clear that there was a beautiful and deep mathematical structure behind the Berry phase \cite{Simon_1983}. Since then, the Berry phase is usually discussed in terms of differential geometry, which makes it clear that the Berry phase is an example of holonomy, a concept well known to professional physicists from general relativity and gauge theories \cite{Nakahara_2003}.

Of course, the approach through differential geometry \cite{Chruscinski_2004,Bohm_2003} provides a much broader perspective on geometric phase than elementary quantum mechanical considerations \cite{Sakurai_2011,griffiths_2018,Commins_2014}. However, for a beginner, the language of fiber bundles may seem too abstract and obscure. For good pedagogy, it is desirable to introduce and illustrate the concept with simple and clear concrete examples. For this purpose, a spin-half system in a rotating magnetic field is usually used \cite{Sakurai_2011,griffiths_2018}. The spin-one system in a rotating magnetic or electric field is another candidate \cite{Budker_2008,Vutha_2009,Zhang_2006,Xu_2006}.

In \cite{Budker_2008} the Berry phase is introduced through considering spin-one system in a rotating electric field. According to our observations, students find it quite challenging to understand the solution of this problem as it is given in the book. Therefore, in this note we decided to provide a more detailed solution. Throughout the article, Einstein's summation convention and atomic units $\hbar=1$ are assumed (if not otherwise stated).

\section{Energy levels in a stationary electric field}
\label{ch2}
Let us consider a system with spin one in a constant electric field $\vec{\mathbfcal E}$ pointing in the $z$ direction. In the absence of the field, the energy levels of the states $|1,1\rangle$, $|1,0\rangle$ and $|1,-1\rangle$ are degenerate due to the assumed rotational symmetry. The interaction Hamiltonian has the form $\hat{H}_{int}=-\hat{\vec{D}}\cdot\vec{\mathbfcal E}=-\hat{D}_z{\cal E}$, where  $\hat{\vec{D}}$ is the electric dipole moment operator. In atomic electric dipole transitions induced by $\hat{D}_z$, the selection rules are $\Delta m=0$, $\Delta l=\pm 1$ \cite{griffiths_2018}. Therefore, we assume that $\hat{H}_{int}$ cannot mix states $|1,m\rangle$, $m=1,0,-1$, among themselves. With our simplified example, we want to simulate behaviour of an actual experimentally observed systems that has an angular momentum equal to one, and is subjected to an electric field. Such system exhibits coupling between $|1,0\rangle$ and  $|0,0\rangle$ states. Indeed, this is not forbidden theoretically by   $\Delta l = \pm1$ selection rules, and can be seen experimentally as electric dipole transitions.  So, in order to mimic this $\Delta l = \pm1$ selection rule when calculating the effect of electric field, we do have to extend the system to include the $|0,0\rangle$ state \cite{Vutha_2009}. 

Since our goal is simply to illustrate the Berry phase, we will consider this simple four-state system and not a real atom with total angular momentum $F=1$. However, we assume that $\hat{H}_{int}$ couples only $|1,0\rangle$ and $|0,0\rangle$ states. 

It is important to notice that $\langle l,m |l^{\prime},m^{\prime}\rangle = \delta_{ll^{\prime}}\delta_{mm^{\prime}}$, which means that the system of angular states $|1,1\rangle$, $|1,0\rangle$, $|1,-1\rangle$, and $|0,0\rangle$ would represent an orthonormal basis. One possible proof of this is done by acknowledging that $\langle \theta, \varphi| l,m\rangle = Y_{l,m}(\theta, \varphi)$, i. e. $| l,m\rangle$ are spherical harmonics in $\theta$, $\varphi$ coordinate representation, and the spherical harmonics are orthonormal.

Now we diagonalize the total Hamiltonian $\hat H=\hat H_0+\hat H_{int}$. It is already diagonal in the subspace spanned by the states $|1,1\rangle$ and $|1,-1\rangle$. If $\langle 1,0|\hat{D}_z|0,0\rangle=\langle 0,0|\hat{D}_z|1,0\rangle=d$, where $d$ is some real number, then in the orthonormal basis of the $|1,0\rangle$ and $|0,0\rangle$ states the remaining part of $\hat H$ is given by the following matrix
\begin{equation}
H=\left (\begin{array}{cc} E_1 & -d{\cal E} \\ -d{\cal E} & -E_1\end{array}\right ),\;\;|1,0\rangle=\left (\begin{array}{c} 1\\ 0\end{array}\right ),\;\; |0,0\rangle=\left (\begin{array}{c} 0\\ 1\end{array}\right ),
\label{eq1}
\end{equation}
where $E_1$ and $-E_1$ are the unperturbed energies (in the absence of an electric field) of the states $|1,0\rangle$ and $|0,0\rangle$ (it is convenient to choose the zero for energy halfway between unperturbed $|1,0\rangle$ and $|0,0\rangle$ states \cite{Vutha_2009}).

The Hamiltonian (\ref{eq1}) is diagonalized by the following transformation
\begin{equation}
|0\rangle=\cos{\alpha} |1,0\rangle+\sin{\alpha} |0,0\rangle, \;\;\;
|00\rangle=-\sin{\alpha} |1,0\rangle+\cos{\alpha} |0,0\rangle,
\label{eq2}
\end{equation}
with
\begin{equation}
\tan{2\alpha}=-\frac{d{\cal E}}{E_1},\;\;\cos{2\alpha}=\frac{E_1}{\sqrt{E_1^2+d^2{\cal E}^2}},\;\; \sin{2\alpha}=-\frac{d{\cal E}}{\sqrt{E_1^2+d^2{\cal E}^2}}.
\label{eq3}
\end{equation}
In addition, the energy eigenvalues are
\begin{equation}
E_0=\cos{2\alpha}\,E_1-\sin{2\alpha}\,d{\cal E}=\sqrt{E_1^2+d^2{\cal E}^2},\;E_{00}=-\sqrt{E_1^2+d^2{\cal E}^2}.
\label{eq4}
\end{equation}
Note that the $|0\rangle$ and $|00\rangle$ states no longer have a definite total spin, but the spin's $z$-projection remains a good quantum number (the electric field breaks rotational symmetry, but rotation around the field direction remains a symmetry).

If $d{\cal E}\ll E_1$, then the angle $\alpha$ is small (and negative). In this case, $|0\rangle$ is mainly the state $|1,0\rangle$, and $|00\rangle$ is mainly the state $|0,0\rangle$.

\section{Rotations: active and passive}
In physics the word ``Rotation" can have two distinct meanings, and one can pretty rapidly become lost in confusion if one doesn't properly keep track of what is rotated against what \cite{Chaichian_1998,Morrison_1987,Millot_2012}. An active rotation is a rotation of a vector in a fixed coordinate system. An active rotation produces a new vector. Passive rotation is the rotation of the coordinate system without changing the vector. In addition, understanding that the unit vectors of the rotated coordinate system are produced by executing an active transformation on the unit vectors of the original coordinate system may be useful.

Let $\vec{e}_i$ be unit vectors along coordinate axes in ${\mathbb R}^3$. Then, under an active rotation $\hat{R}_n^{(a)}(\varphi)$ around some axis $\vec{e}_n$, the vector $\vec{r}=x_{i}\vec{e}_i$ changes to the vector $\vec{r}^{\prime}=x_{i}\hat{R}_n^{(a)}(\varphi)\vec{e}_i$.
The matrix $R=(R_{ij})$ associated with the rotation $\hat{R}_n^{(a)}(\varphi)$ can be introduced by expanding $\hat{R}_n^{(a)}( \varphi)\vec{e}_i=R_{ji}\vec{e}_j$. Then the expression for $\vec{r}^{\prime}$ above transforms into $\vec{r}^{\prime}=x_{i}R_{ji}\vec{e}_j$, at the same time $\vec{r}^{\prime}$ can be written in term of its own new coordinates: $\vec{r}^{\prime} =  x^{\prime}_{j}\vec{e}_j$. By combining these two different expressions for $\vec{r}^{\prime}$ one can see that 
$x^{\prime}_{j} = R_{ji}x_{i}$.

Under a passive rotation $\hat{R}_n^{(p)}(\varphi)$ the coordinates of the vector  $\vec{r}=x_{i}\vec{e}_i$ change to $x_i^\prime=\vec{r}\cdot \hat{R}_n^{(a)}(\varphi)\vec{e}_i=x_j\vec{e}_j\cdot \hat{R}_n^{(a)}(\varphi)\vec{e}_i=R_{ki}x_j\vec{e}_j\cdot\vec{e}_k=R_{ji}x_j=R^T_{ij}x_j$. Therefore, the matrix which corresponds to $\hat{R}_n^{(p)}(\varphi)$ is the transpose of the matrix which corresponds to $\hat{R}_n^{(a)}(\varphi)$. Since the rotation matrices must be orthogonal so that the lengths of the vectors remain the same, this means that all points will have exactly the same coordinates relative to the new coordinate system as before the rotations, if you first rotate the coordinate system, and then the radius vectors of these points by the same amount (or in the reverse order), and this makes perfect sense, since after these operations the relative proximity of points and axes of the coordinate system does not change.

Operator $\hat{R}_3^{(a)}(\varphi)$ is an operator of rotation by angle $\varphi$ around axis $\vec{e}_3$, in our case of orthonormal basis $(\vec{e}_1, \vec{e}_2, \vec{e}_3)$, this is the same as rotation by angle $\varphi$ in a plane spanned by ($\vec{e}_1, \vec{e}_2$), so $\hat{R}_3^{(a)}(\varphi)\vec{e}_1=\cos{\varphi}\,\vec{e}_1+\sin{\varphi}\,\vec{e}_2$ and $\hat{R}_3^{(a)}(\varphi)\vec{e}_2=-\sin{\varphi}\,\vec{e}_1+\cos{\varphi}\,\vec{e}_2$. Therefore, under passive rotation $\hat{R}_3^{(p)}(\varphi)$ we will have
\begin{equation}
x^\prime_1=\cos{\varphi}\,x_1+\sin{\varphi}\,x_2,\;\;\;  x^\prime_2=-\sin{\varphi}\,x_1+\cos{\varphi}\,x_2,\;\;\;x^\prime_3=x_3,
\label{eq5}
\end{equation}
and these relations remain valid under cyclic permutation of $x_1,x_2,x_3$. Hence the rotation matrices which correspond to passive rotations $\hat{R}_3^{(p)}(\varphi)$ and $\hat{R}_2^{(p)}(\varphi)$  are
\begin{equation}
R_3(\varphi)=\left (\begin{array}{ccc} \cos{\varphi} & \sin{\varphi} & 0 \\ -\sin{\varphi} & \cos{\varphi} & 0 \\ 0 & 0 & 1\end{array}\right ),\;\;
R_2(\varphi)=\left (\begin{array}{ccc} \cos{\varphi} & 0 & -\sin{\varphi} \\ 0 & 1 & 0 \\ \sin{\varphi} & 0 & \cos{\varphi} \end{array}\right ).   
\label{eq6}
\end{equation}
Consider a rotating electric field
\begin{equation}
\vec{\mathbfcal E}={\cal E}\left (\sin{\theta}\cos{\varphi(t)}\vec{e}_1+\sin{\theta}\sin{\varphi(t)}\vec{e}_2+
\cos{\theta}\vec{e}_3\right ).
\label{eq7}
\end{equation}
If we rotate the coordinate system first by the angle $\varphi(t)$ around the $z$ axis, and then by the angle $\theta$ around the new $y$ axis, the electric field ${\cal E } $ will be along the third axis and will not rotate in the new coordinate system. The rotation operator $\hat R^{(a)}=\hat{R}^{(a)}_{2^\prime}(\theta)\hat{R}^{(a)}_3(\varphi)$ mixes the active and passive viewpoints since $\hat{R}^{(a)}_{2^\prime}(\theta)$ is the rotation around rotated axis $y^\prime$. It is useful to rewrite $\hat R$ solely in terms of rotations around fixed axes. This is possible thanks to the 
so-called class angle relation \cite{Morrison_1987,biedenharn_1984}: if the axis $\vec{n}^\prime$ is obtained from the axis $\vec{n}$ by rotation $\hat G^{(a)}$ then $\hat{R}^{(a)}_{n^\prime}(\varphi)=\hat{G}^{(a)}\hat{R}^{(a)}_n(\varphi)[\hat{G}^{(a)}]^{-1}$. Therefore,
\begin{equation}
\hspace*{-15mm}
\hat R^{(a)}= \hat{R}^{(a)}_{2^\prime}(\theta)\hat{R}^{(a)}_3(\varphi)= \hat{R}^{(a)}_3(\varphi)\hat{R}^{(a)}_2(\theta)[\hat{R}^{(a)}_3(\varphi)]^{-1}\hat{R}^{(a)}_3(\varphi)=\hat{R}^{(a)}_3(\varphi)\hat{R}^{(a)}_2(\theta). 
\label{eq8}
\end{equation}
Notice that in our case of rotating electric field according to (\ref{eq7}), $\theta$ angle is constant, but $\varphi$ angle depends on time, so this operator depends on time too. Explicit form of the corresponding matrix, considering $\hat R$ as a passive rotation, can be obtained by using (\ref{eq6}) and remembering that the matrix which corresponds to $\hat{R}_n^{(p)}(\varphi)$ is the transpose of the matrix which corresponds to $\hat{R}_n^{(a)}(\varphi)$:
\begin{equation}
\hspace*{-15mm}
R= \left [R_3^T(\varphi)R_2^T(\theta)\right ]^T=R_2(\theta)R_3(\varphi)=\left(\begin{array}{ccc} \cos{\theta}\cos{\varphi} & \cos{\theta}\sin{\varphi} & -\sin{\theta} \\ -\sin{\varphi} & \cos{\varphi} & 0 \\ \sin{\theta}\cos{\varphi} & \sin{\theta}\sin{\varphi} & \cos{\theta} \end{array} \right ).
\label{eq9}
\end{equation}
Generators of rotation ($\hat{J}_1, \hat{J}_2, \hat{J}_3$) correspondingly around ($\vec{e}_1, \vec{e}_2, \vec{e}_3$) are defined in the usual way \cite{Morrison_1987}: $\hat{R}^{(a)}_3(\varphi)=e^{-i\varphi\hat{J}_3}$, $\hat{R}^{(a)}_2(\varphi)=e^{-i\varphi\hat{J}_2}$. If $\vec{r}=x_i\vec{e}_i$, then by plugging $\Delta \varphi\ll 1$ into transposed \ref{eq6} (because active rotation matrix equals to transposed passive), and using the fact that $\sin(\Delta \varphi)\approx \Delta \varphi$, $\cos(\Delta \varphi)\approx 1$, one can see that the rotated vector $\vec{r}^{\prime}=\hat{R}^{(a)}_3(\Delta \varphi)\vec{r}\approx x_1(\vec{e}_1+\Delta \varphi \vec{e}_2)+x_2(-\Delta \varphi \vec{e}_1+\vec{e}_2)+x_3\vec{e}_3$. 

The same vector $\vec{r}^{\prime}=\hat{R}^{(a)}_3(\Delta \varphi)\vec{r} = e^{-i\Delta \varphi\hat{J}_3}\vec{r}$ by definition of generator $\hat{J}_3$. By expanding the exponent $e^{-i\Delta \varphi\hat{J}_3} \approx 1 -i\Delta \varphi\hat{J}_3$ one can see that $\vec{r}^{\prime} \approx  \vec{r}-i\Delta\varphi\left (x_1\hat{J}_3\vec{e}_1+x_2\hat{J}_3\vec{e}_2+x_3\hat{J}_3\vec{e}_3\right ) $.
The comparison of two different expressions written above for $\vec{r}^{\prime}$, and knowing that $(x_1,x_2,x_3)$ might be arbitrary, yields the following results 
\begin{equation}
\hat{J}_3\vec{e}_1=i\vec{e}_2,\;\; \hat{J}_3\vec{e}_2=-i\vec{e}_1,
\label{eq10}
\end{equation}
and because generators are linear operators
\begin{equation}
\hat{J}_3\left (\vec{e}_1+i\vec{e}_2\right )= \vec{e}_1+i\vec{e}_2,\;\;\hat{J}_3\left (\vec{e}_1-i\vec{e}_2\right )=-\left(\vec{e}_1-i\vec{e}_2 \right).
\label{eq11}
\end{equation}
In many applications, instead of the Cartesian basis $\vec{e}_i$, $i=1,2,3$, it is more convenient to use the spherical basis \cite{Budker_2008,Devanathan_2002}
\begin{equation}
\vec{\epsilon}_1=-\frac{1}{\sqrt{2}}\left (\vec{e}_1+i\vec{e}_2\right ),\;\;\vec{\epsilon}_0=\vec{e}_3,\;\;\vec{\epsilon}_{-1}=\frac{1}{\sqrt{2}}\left (\vec{e}_1-i\vec{e}_2\right ).
\label{eq12}
\end{equation}
Then, for example, the components of the radius-vector in the spherical basis are
\begin{equation}
r_1=-\frac{1}{\sqrt{2}}\left (x_1+ix_2\right ),\;\;r_0=x_3,\;\;r_{-1}=\frac{1}{\sqrt{2}}\left (x_1-ix_2\right ).
\label{eq13}
\end{equation}
According to (\ref{eq12}), $\vec{\epsilon}_i$, $i=1,0,-1$, are eigenvectors of the operator $\hat{J}_3$, and the coefficients in (\ref{eq12}) are chosen so that the following relations hold ($m,n=1,0,-1$)
\begin{equation}
\vec{\epsilon}^{\;*}_m=(-1)^m\vec{\epsilon}_{-m},\;\;\vec{\epsilon}^{\;*}_m\cdot \vec{\epsilon}_n=\delta_{mn},\;\;\vec{r}=r_m\vec{\epsilon}^{\;*}_m=\sum\limits_{m=1,0,-1}(-1)^mr_m\vec{\epsilon}_{-m}.
\label{eq14}
\end{equation}
Using $x_1=\frac{1}{\sqrt{2}}\left (r_{-1}-r_1\right )$, $x_2=\frac{i}{\sqrt{2}}\left (r_{-1}+r_1\right )$, $x_3=r_0$, it is straightforward to rewrite the rotation matrices (\ref{eq6}) so that the spherical components replace the Cartesian ones \cite{Devanathan_2002}:
\begin{equation} 
\hspace*{-25mm}
{\cal R}_3(\varphi)=\left (\begin{array}{ccc} e^{-i\varphi} & 0 & 0 \\ 0 & 1 & 0 \\ 0 & 0 &  e^{i\varphi}\end{array}\right ),\; {\cal R}_2(\varphi)=\left (\begin{array}{ccc} \frac{1}{2}(1+\cos{\varphi}) & \frac{1}{\sqrt{2}}\sin{\varphi} &  \frac{1}{2}(1-\cos{\varphi})\\ - \frac{1}{\sqrt{2}}\sin{\varphi} & \cos{\varphi} & \frac{1}{\sqrt{2}}\sin{\varphi} \\ \frac{1}{2}(1-\cos{\varphi}) & - \frac{1}{\sqrt{2}}\sin{\varphi}  &  \frac{1}{2}(1+\cos{\varphi}) \end{array}\right ).
\label{eq15}
\end{equation}
However, since the radius-vector $\vec{r}=r^*_m\epsilon_m$, we can also use
$$\left (\begin{array}{c} r^*_1 \\ r^*_0 \\ r^*_{-1}\end{array}\right )= \left (\begin{array}{c} -r_{-1} \\ r_0 \\ -r_1\end{array}\right ),\;\;\;\mathrm{instead \;\;of}\;\;\; \left (\begin{array}{c} r_1 \\ r_0 \\ r_{-1}\end{array}\right ),$$
as its coordinate representation. In this case, ${\cal R}_3(\varphi)$ will be replaced by its complex conjugate matrix, while ${\cal R}_2(\varphi)$ will remain unchanged (because it is a real matrix), and
\begin{equation}
\hspace*{-10mm}
{\cal R}= {\cal R}_2(\theta) {\cal R}^*_3(\varphi)= \left (\begin{array}{ccc} \frac{1}{2}(1+\cos{\theta})e^{i\varphi} & \frac{1}{\sqrt{2}}\sin{\theta} &  \frac{1}{2}(1-\cos{\theta})e^{-i\varphi}\\ -\frac{1}{\sqrt{2}}\sin{\theta}e^{i\varphi} & \cos{\theta} & \frac{1}{\sqrt{2}}\sin{\theta}e^{-i\varphi} \\ \frac{1}{2}(1-\cos{\theta})e^{i\varphi} & -\frac{1}{\sqrt{2}}\sin{\theta}  &  \frac{1}{2}(1+\cos{\theta})e^{-i\varphi} \end{array}\right ). 
\label{eq16} 
\end{equation}

\section{Rotations of the quantum mechanical states}
In a given reference frame, we use $|1,1\rangle$,  $|1,0\rangle$, $|1,-1\rangle$ and  $|0,0\rangle$ as basic states for our quantum system. 
The notation $|l, m\rangle$ that we employ signifies that $l$ represents the angular momentum quantum number, while $m$ represents the angular momentum projection quantum number. In the rotating frame, where the electric field stays still and aligned with the z axis, according to Section \ref{ch2}, eigenstates of the Hamiltonian have the following components
\begin{equation}
\hspace*{-10mm}
\phi_1=\left (\begin{array}{c} 1 \\ 0 \\ 0 \\ 0\end{array}\right),\;\;\phi_0=\left (\begin{array}{c} 0 \\ \cos{\alpha} \\ 0 \\ \sin{\alpha}\end{array}\right),\;\;\phi_{-1}=\left (\begin{array}{c} 0 \\ 0 \\ 1 \\ 0\end{array}\right),\;\; \phi_{00}=\left (\begin{array}{c} 0 \\ -\sin{\alpha} \\ 0 \\ \cos{\alpha} \end{array}\right).
\label{eq17}
\end{equation}
When the reference system is changed, the basis elements $|1,m\rangle$ will be transformed
\begin{equation}
|1,m\rangle^\prime=\hat{R}^{(a)}|1,m\rangle=\langle 1,n|\hat{R}^{(a)}|1,m\rangle |1,n\rangle,
\label{eq18}
\end{equation}
while $|0,0\rangle$ and the quantum state $\Psi$ itself will remain unchanged. Therefore, the elements of the column-wave function $\Psi$ will transform as follows: $$\Psi^\prime_{00}=\langle0,0|^\prime\Psi\rangle=\langle 0,0|\Psi\rangle=\Psi_{00},$$ while
\begin{equation}
\hspace*{-15mm} \Psi^\prime_{m}=\langle1,m|^\prime\Psi\rangle=\langle 1,m|\hat{R}^{(a)\,+}|\Psi\rangle=\langle 1,m|\hat{R}^{(a)\,+}|1,n\rangle\langle 1,n|\Psi\rangle={\cal R}_{mn}\Psi_n,
\label{eq19}
\end{equation}
where $\langle 1,m|^\prime=(|1,m\rangle^\prime)^+$ and
\begin{equation} 
\hspace*{-15mm}
{\cal R}_{mn}=\langle 1,m|\hat{R}^{(a)\,+}|1,n\rangle =  \langle 1,m| e^{i\theta \hat{J}_2}e^{i\varphi \hat{J}_3}|1,n\rangle= e^{in\varphi} \langle 1,m| e^{i\theta \hat{J}_2}|1,n\rangle.
\label{eq20}
\end{equation}
But $\langle 1,m| e^{i\theta \hat{J}_2}|1,n\rangle=d^1_{mn}(-\theta)$, with $d^1_{mn}(\theta)=\langle 1,m| e^{-i\theta \hat{J}_2}|1,n\rangle$ as the Wigner rotation matrix. Using \cite{Morrison_1987}
\begin{equation}
d^1(\theta)=\left (\begin{array}{ccc}   \frac{1}{2}(1+\cos{\theta}) & -\frac{\sin{\theta}}{\sqrt{2}} & \frac{1}{2}(1-\cos{\theta}) \\  \frac{\sin{\theta}}{\sqrt{2}} & \cos{\theta} & -\frac{\sin{\theta}}{\sqrt{2}} \\ \frac{1}{2}(1-\cos{\theta}) & \frac{\sin{\theta}}{\sqrt{2}} & \frac{1}{2}(1+\cos{\theta}) \end{array}\right ),
\label{eq21}
\end{equation}
it can be easily checked that $(d^1_{mn}(-\theta)e^{in\varphi})$, $m,n=1,0,-1$, is indeed the matrix ${\cal R}$ from (\ref{eq16}).

It follows from the above that the eigenstates of the Hamiltonian in the rotating and initial (laboratory) coordinate systems are related by the relation $\phi_m={\cal R}\psi_n$, $n=1,0,-1$. Consequently, the eigenstates in the laboratory frame are $\psi_n={\cal R}^+\phi_n$, since the matrix ${\cal R}$ is unitary. To extend this relation to the $n=00$ state, we trivially extend the ${\cal R}^+$ matrix using the fact that the $\langle 0,0|$ state is invariant under rotations:
\begin{equation}
{\cal R}^+=\left (\begin{array}{cccc} \frac{1}{2}(1+\cos{\theta})e^{-i\varphi} & -\frac{1}{\sqrt{2}}\sin{\theta}e^{-i\varphi} &  \frac{1}{2}(1-\cos{\theta})e^{-i\varphi} & 0\\ \frac{1}{\sqrt{2}}\sin{\theta} & \cos{\theta} & -\frac{1}{\sqrt{2}}\sin{\theta} & 0 \\ \frac{1}{2}(1-\cos{\theta})e^{i\varphi} & \frac{1}{\sqrt{2}}\sin{\theta}e^{i\varphi}  &  \frac{1}{2}(1+\cos{\theta})e^{i\varphi} & 0 \\ 0 & 0 & 0 & 1 \end{array}\right ). 
\label{eq22}
\end{equation} 
Notice that $\cal R^+$
depends on angles $\varphi$ and $\theta$, and only the angle $\varphi$ of rotating system depends on time, so $\cal R^+$ is time dependent too.
Using (\ref{eq17}) and  (\ref{eq22}), we find explicit forms of the eigenstates $\psi_n={\cal R}^+\phi_n$ in the laboratory frame 
\begin{eqnarray} 
\psi_1=\left (\begin{array}{c}  \frac{1}{2}(1+\cos{\theta})e^{-i\varphi} \\  \frac{1}{\sqrt{2}}\sin{\theta} \\  \frac{1}{2}(1-\cos{\theta})e^{i\varphi} \\ 0\end{array}\right)&,&\;\;\;
\psi_0=\left (\begin{array}{c} -\frac{1}{\sqrt{2}}\sin{\theta}e^{-i\varphi}\cos{\alpha}  \\  \cos{\theta}\cos{\alpha} \\  \frac{1}{\sqrt{2}}\sin{\theta}e^{i\varphi}\cos{\alpha}  \\ \sin{\alpha} \end{array}\right),
\nonumber \\ 
\psi_{-1}=\left (\begin{array}{c} \frac{1}{2}(1-\cos{\theta})e^{-i\varphi} \\  -\frac{1}{\sqrt{2}}\sin{\theta} \\  \frac{1}{2}(1+\cos{\theta})e^{i\varphi} \\ 0\end{array}\right)&,&\;\;\;
\psi_{00}=\left (\begin{array}{c} \frac{1}{\sqrt{2}}\sin{\theta}e^{-i\varphi}\sin{\alpha}  \\  -\cos{\theta}\sin{\alpha} \\  -\frac{1}{\sqrt{2}}\sin{\theta}e^{i\varphi}\sin{\alpha}  \\ \cos{\alpha} \end{array}\right).
\label{eq23}
\end{eqnarray}

\section{Calculation of the Berry phase}
Let's recall that $\phi_n$ is  not dependent on time being an eigenstate of the Hamiltonian in the rotating system, where the electric field stays still and aligned with the z axis. At some given moment $t$, 
$\psi_n(t)={\cal R}^+(t)\phi_n$ is an eigenstate of the Hamiltonian in the laboratory system. Since $\psi_n(t)={\cal R}^+(t)\phi_n$ is an eigenstate of the Hamiltonian, naively one can think that its development in time is described by the usual phase factor $e^{-iE_nt}$. However, what hinders this is the dependence of the rotation matrix ${\cal R}^+$ on time. To find the correct time-dependent eigenstate $\Psi_n(t)$, which by definition is a solution of the time-dependent Schr\"{o}dinger equation 
\begin{equation}
i\frac{\partial \Psi_n}{\partial t}=\hat H \Psi_n=E_n\Psi_n,
\label{eq25}
\end{equation}
we use a common mathematical physics technique. We introduce an unknown functional multiplier $c_n(t)$ which turns out to be the phase factor, to get the ansatz  
\begin{equation}
\Psi_n(t)=c_n(t)e^{-iE_nt}\psi_n(t).
\label{eq24}
\end{equation}
Substituting (\ref{eq24}) into (\ref{eq25}), we get the condition
\begin{equation}
\frac{dc_n}{dt}\psi_n(t)+c_n\frac{d\psi_n}{dt}=0,
\label{eq26}
\end{equation}
which, after multiplying from the left by $\psi_n^+(t)$, takes the form
\begin{equation}
\frac{dc_n}{dt}=-\psi_n^+\frac{d\psi_n^+}{dt}\,c_n.
\label{eq27}
\end{equation}
Using the explicit expressions (\ref{eq23}), we can calculate
\begin{equation}
\hspace{-15mm}
\psi^+_1\frac{d\psi_1}{dt}=-i\frac{d\varphi}{dt}\cos{\theta},\;\;  \psi^+_0\frac{d\psi_0}{dt}=0,\;\; \psi^+_{-1}\frac{d\psi_{-1}}{dt}=i\frac{d\varphi}{dt}\cos{\theta},\;\; \psi^+_{00}\frac{d\psi_{00}}{dt}=0.
\label{eq29}
\end{equation}
Combined with (\ref{eq27}), these results imply (we assume that multiplying by $00$ is the same as multiplying by zero)
\begin{equation}
c_n(t)=c_n(0)e^{in\varphi(t)\cos{\theta}}.
\label{eq30}
\end{equation}
Therefore, after a full period $T$ of revolution of the electric field, $\varphi(T)=2\pi$ and the wave function $\Psi_n$ acquires an additional geometric phase factor $e^{i\alpha_g}$ with $ \alpha_g=2\pi n\,cos{\theta}$. Since the phase is determined only up to an integer multiple of $2\pi$, and since $\Omega=2\pi (1-\cos{\theta})$ is the solid angle subtended by the electric field vector over one cycle, the Berry phase $\alpha_g$ can be rewritten as
\begin{equation}
\alpha_g=-n\Omega.
\label{eq31}
\end{equation}
We don't actually need explicit forms (\ref{eq23}) to calculate $\alpha_g$. Indeed, in light of (\ref{eq17})
\begin{eqnarray} &&
\hspace{-20mm}
\phi_1^+{\cal R}\frac{d{\cal R}^+}{dt}\phi_1=\langle 1,1|\hat{R}^{(p)}\frac{d\hat{R}^{(p)+}}{dt}|1,1\rangle,\;\;\;
\phi_{-1}^+{\cal R}\frac{d{\cal R}^+}{dt}\phi_{-1}=\langle 1,-1|\hat{R}^{(p)}\frac{d\hat{R}^{(p)+}}{dt}|1,-1\rangle, \nonumber \\ && \hspace{-20mm}
\phi_0^+{\cal R}\frac{d{\cal R}^+}{dt}\phi_0=\cos^2{\alpha}\,\langle 1,0|\hat{R}^{(p)}\frac{d\hat{R}^{(p)+}}{dt}|1,0\rangle+\sin^2{\alpha}\,\langle 0,0|\hat{R}^{(p)}\frac{d\hat{R}^{(p)+}}{dt}|0,0\rangle \nonumber \\ && \hspace{-20mm}+\sin{\alpha}\cos{\alpha}\left (\langle 1,0|\hat{R}^{(p)}\frac{d\hat{R}^{(p)+}}{dt}|0,0\rangle+\langle 0,0|\hat{R}^{(p)}\frac{d\hat{R}^{(p)+}}{dt}|1,0\rangle\right ), \nonumber \\ && \hspace{-20mm}
\phi_{00}^+{\cal R}\frac{d{\cal R}^+}{dt}\phi_{00}=\sin^2{\alpha}\,\langle 1,0|\hat{R}^{(p)}\frac{d\hat{R}^{(p)}+}{dt}|1,0\rangle+\cos^2{\alpha}\,\langle 0,0|\hat R^{(p)}\frac{d\hat{R}^{(p)+}}{dt}|0,0\rangle \nonumber \\ && \hspace{-20mm} -\sin{\alpha}\cos{\alpha}\left (\langle 1,0|\hat{R}^{(p)}\frac{d\hat{R}^{(p)+}}{dt}|0,0\rangle+\langle 0,0|\hat{R}^{(p)}\frac{d\hat{R}^{(p)+}}{dt}|1,0\rangle\right ),
\label{eq32}
\end{eqnarray}
where, according to (\ref{eq20}), $\hat{R}^{(p)}=e^{i\theta\hat{J}_2}e^{i\varphi\hat{J}_3}=\hat{R}^{(a)+}$, and consequently
\begin{equation}
\hat{R}^{(p)}\frac{d\hat{R}^{(p)+}}{dt}=-i\frac{d\varphi}{dt}e^{i\theta\hat{J}_2}e^{i\varphi\hat{J}_3}\hat{J}_3e^{-i\varphi\hat{J}_3}e^{-i\theta\hat{J}_2}=
-i\frac{d\varphi}{dt}e^{i\theta\hat{J}_2}\hat{J}_3e^{-i\theta\hat{J}_2}.
\label{eq33}
\end{equation}
We can calculate (\ref{eq33}) using one of the Baker-Campbell-Hausdorff formulas \cite{Bjorken_1964,Achilles_2012}
\begin{equation}
\hspace*{-20mm}
e^{i\theta\hat{J}_2}\hat{J}_3e^{-i\theta\hat{J}_2}=\hat{J}_3+i\theta[\hat{J}_2,\hat{J}_3]+\frac{(i\theta)^2}{2!}[\hat{J}_2,[\hat{J}_2,\hat{J}_3]]+ \frac{(i\theta)^3}{3!}[\hat{J}_2,[\hat{J}_2,[\hat{J}_2,\hat{J}_3]]]+\ldots
\label{eq34}
\end{equation}
But, according to commutation relations $[\hat{J}_i,\hat{J}_j]=i\epsilon_{ijk}\hat{J}_k$, we have $[\hat{J}_2,\hat{J}_3]=i\hat{J}_1$, $[\hat{J}_2,[\hat{J}_2,\hat{J}_3]]=\hat{J}_3$, and consequently (\ref{eq34}) takes the form
\begin{equation}
\hspace*{-15mm}
e^{i\theta\hat{J}_2}\hat{J}_3e^{-i\theta\hat{J}_2}=\hat{J}_3\sum\limits_{n=0}^\infty\frac{(-1)^n\theta^{2n}}{(2n)!}-\hat{J}_1\sum\limits_{n=0}^\infty\frac{(-1)^n\theta^{2n+1}}{(2n+1)!}= 
\hat{J}_3\cos{\theta}- \hat{J}_1\sin{\theta}.
\label{eq35}
\end{equation}
Since $|0,0\rangle$ is annihilated by any of $\hat{J}_i$ generators, and since $\hat{J}_1$ does not have non-zero diagonal matrix elements in the subspace spanned by $|1,1\rangle$, $|1,0\rangle$, $|1,-1\rangle$, we get from
(\ref{eq32}) and (\ref{eq35})
\begin{equation}
\phi_n^+{\cal R}\frac{d{\cal R}^+}{dt}\phi_n=-in\frac{d\varphi}{dt}\cos{\theta},
\label{eq36}
\end{equation}
the result equivalent to (\ref{eq29}).

\section{What is the role of the adiabatic approximation?}
In the previous section, we treated the adiabatic approximation somewhat casually, sacrificing rigor for the sake of simplicity. Now it's time to fix some inconsistencies.

Actually (\ref{eq26}) is true only in the extreme adiabatic limit, which is incompatible with cyclic evolution. Any cyclic evolution assumes a finite rate of change, and therefore (\ref{eq24}) will eventually have a small admixture of other eigenstates (controlled by the $\epsilon_m$ parameter) \cite{griffiths_2018}:
\begin{equation}
\Psi_n(t)=c_n(t)e^{-iE_nt}{\cal R}^+(t)\phi_n+\sum\limits_{m\ne n}\epsilon_m c_m(t)e^{-iE_mt}{\cal R}^+(t)\phi_m,
\label{eq37}
\end{equation}
and after substituting this into the Schrödinger equation, we get instead of (\ref{eq26})
\begin{equation}
\hspace*{-15mm}
\frac{dc_n}{dt}{\cal R}^+\phi_n+c_n\frac{d{\cal R}^+}{dt}\phi_n=-\sum\limits_{m\ne n}\epsilon_me^{-i(E_m-E_n)}\left ( \frac{dc_m}{dt}{\cal R}^+\phi_m+c_m\frac{d{\cal R}^+}{dt}\phi_m\right ).
\label{eq38}
\end{equation}
When we multiply both sides of this equation from the left by $\phi_n^+{\cal R}$, the first term in the right-hand-side of (\ref{eq38}) disappears thanks to the orthogonality of $\phi_n$ and $\phi_m$ when $n\ne m$, but the second term, in general, will survive. As a result, we get
\begin{equation}
\frac{dc_n}{dt}=-\phi_n^+{\cal R}\frac{d{\cal R}^+}{dt}\phi_n\,c_n-\sum\limits_{m\ne n}\epsilon_me^{-i(E_m-E_n)}\,\phi_n^+{\cal R}\frac{d{\cal R}^+}{dt}\phi_m\,c_m.
\label{eq39}
\end{equation}
If $E_m\ne E_n$, then under adiabatic evolution $\epsilon_m\ll 1$. Consequently the second term in (\ref{eq39}) is of the second order (since $\frac{d{\cal R}^+}{dt}\sim\frac{d\varphi}{dt}\ll 1$) and can be neglected. In fact, in this case, the parameter characterizing the departure from the adiabaticity $\epsilon_m\sim T_i/T_e$, where $T_i$ is the characteristic time scale of the internal dynamics of the system under consideration, and $T_e$ is the external time scale over which the system parameters significantly change \cite{griffiths_2018}. In our case, $T_i\sim\frac{1}{E_0-E_1}$ and $T_e\sim 1/\frac{d\varphi}{dt}$, so that the adiabaticity condition is (we temporarily restored $\hbar$) $\hbar\frac{d\varphi}{dt}\ll E_0-E_1$.

There is an additional complication due to the degeneracy $E_1=E_{-1}$. Then $E_m=E_n$, $\epsilon_m$ can be large and the above reasoning is inapplicable. Fortunately, however, (\ref{eq23}) implies that
\begin{equation}
\psi_1^+\frac{d\psi_{-1}}{dt}=\psi_{-1}^+\frac{d\psi_1}{dt}=0.
\label{eq40}
\end{equation}
Alternatively, we can get the same result if we use (\ref{eq35}) and remember that neither $\hat{J}_3$ nor $\hat{J}_1$ connect $|1,1\rangle$ and $|1,-1\rangle$ states. As we can see, even if (\ref{eq26}) was not strictly speaking correct, (\ref{eq27}) holds up to more thorough scrutiny.

\section{Why can't the Berry phase be removed?}
A quantum mechanical state is given by a ray in Hilbert space, not by a vector, since the states $|\Psi\rangle$ and $c|\Psi\rangle$ with some complex number $c$ correspond to the same quantum state (if the states are assumed normalized, then $c=e^{i\lambda}$ with some phase $\lambda$). A natural question arises why then it is impossible to remove the Berry phase by simply replacing the representatives of the rays of the Hilbert space (by performing a gauge transformation)
\begin{equation}
|\Psi(t)\rangle^\prime=e^{i\lambda(t)}|\Psi(t)\rangle.
\label{eq41}
\end{equation}
However, the Berry phase is invariant under the gauge transformation (\ref {eq41}): as can be easily seen from (\ref{eq27})
\begin{equation}
\alpha^\prime_g=i\int\limits_0^T\psi_n^{\prime\,+}\,\frac{d\psi^\prime_n}{dt}\, dt=\alpha_g-\int\limits_0^T\frac{d\lambda}{dt} dt=\alpha_g-\lambda(T)+\lambda(0)=\alpha_g,
\label{eq{42}}
\end{equation}
since $\lambda(T)=\lambda(0)$ for cyclic evolution, when we are returning to exactly the same physical state, and we need a unique representative of the corresponding Hilbert space ray to represent that state.

The existence of the Berry phase proves that a quantum system surprisingly retains a ``memory" of its motion in terms of a geometric phase as it undergoes cyclic evolution and returns to its original quantum state \cite{Anandan_1992}. It can be suspected that this is true not only for adiabatic evolution, but also for any cyclic process. Indeed, Aharonov and Anandan removed the adiabaticity constraint and generalized the notion of a geometric phase for non-adiabatic cyclic processes \cite{Anandan_1987}.

\section{Concluding remarks}
In \cite{Budker_2008} the Berry phase is introduced through a simple example of a spin-one system in a rotating electric field. We have added some details to the analysis of \cite{Budker_2008} that we hope will be useful for students who are first exposed to this important concept of quantum physics.

The geometric phase is inherently related to the mathematical structure of quantum mechanics and its proper treatment, as mentioned in the introduction, requires differential geometric language of fiber bundles. Such refinements are outside the scope of this article.

\section*{References}

\bibliography{G_phase}

\end{document}